# Nanolithography and manipulation of graphene using an atomic force microscope

A.J.M. Giesbers[a*], U. Zeitler[a†], S. Neubeck[b], F. Freitag[b], K.S. Novoselov[b], and J.C. Maan[a]

[a]*High Field Magnet Laboratory, Institute for Molecules and Materials, Radboud University Nijmegen, Toernooiveld 7, 6525 ED Nijmegen, The Netherlands.*

[b]*Department of Physics, University of Manchester, M13 9PL, Manchester, UK.*

(05 May 2008)

**Abstract**

We use an atomic force microscope (AFM) to manipulate graphene films on a nanoscopic length scale. By means of local anodic oxidation with an AFM we are able to structure isolating trenches into single-layer and few-layer graphene flakes, opening the possibility of tabletop graphene based device fabrication. Trench sizes of less than 30 nm in width are attainable with this technique. Besides oxidation we also show the influence of mechanical peeling and scratching with an AFM of few layer graphene sheets placed on different substrates.

---

[*] Electronic mail: J.Giesbers@science.ru.nl

[†] Electronic mail: U.Zeitler@science.ru.nl

## 1. Introduction

Carbon is one of the most intensively studied materials in solid state physics. Starting with research activities on graphite six decades ago [1] research on carbon continued to be attractive by the discovery of new carbon allotropes such as Buckminster fullerenes [2], carbon nanotubes [3] and the recent fabrication of isolated single-layers of carbon atoms, graphene [4, 5]. Due to the unique electronic properties and high crystal quality of graphene this discovery triggered a great deal of attention in the following years [6]. Besides its fundamental physical properties, the large charge carrier mobilities of up to 200,000 $cm^2$/Vs [7, 8] (two orders of magnitude larger than silicon MOSFETs) also make graphene a promising candidate for integrated electronic circuitries. Due to its planar geometry it can be integrated rather straightforwardly into the current silicon technology and pave the way for interesting novel nano-electronic devices based on e.g. relativistic p-n junctions [9, 10], size-quantized nano ribbons [11–13] or quantum dots [14, 15].

Currently, most graphene devices are fabricated using state-of the art nanofabrication techniques based on electron beam lithography and subsequent reactive plasma edging. A promising alternative method for the fabrication of proof-of-principle devices may be provided by scanning probe techniques, and, more specifically by AFM-lithography. For traditional semiconductors it was indeed already shown successfully that an atomic force microscope (AFM) can be used to create electronic nanostructures by means of mechanical ploughing [16, 17] or local anodic oxidation [21, 22] providing a table-top method for the fabrication of e.g. quantum point contacts [17, 19], quantum dots [17, 20, 21] and phase coherent quantum rings [22].

In this work we will demonstrate how an AFM can be used to locate and nano-manipulate single and few-layer graphene sheets. We will show that the way in which a graphene sheet can be manipulated depends strongly on the substrate it is placed on. Second, we will demonstrate how graphene sheets can be shaped by means of electrochemical oxidation, an extremely promising technique for desktop proof-of-principle device fabrication.

**2. Experimental techniques, results and discussion**

In the past SPM manipulation techniques have already been shown effective to tear, to fold and unfold, and to oxidize graphitic sheets on highly oriented pyrolytic graphite (HOPG) surfaces [23–28]. These experiments already led to speculations towards its use as a tool for nanofabrication of graphitic devices in general and carbon nanotubes, in particular [29]. An interesting question to address is whether such an approach can also be applied on few-layer graphene placed on a $SiO_2$ substrate. For this means we have deposited graphene flakes on a SIMOX wafer using micromechanical exfoliated natural graphite [4, 5]. Their position and thickness was subsequently determined under an optical microscope and confirmed by AFM imaging. Trying to scratch through or to peel-off single graphene layers from these few-layer flakes was unsuccessful; due to the relatively low sticking force of the graphene to the rather rough surface of $SiO_2$ it was only possible to move or crumble entire flakes in a rather uncontrolled fashion.

In order to make the graphene stick better to the substrate we placed it on a flat, epi-ready GaAs substrate using the same exfoliation technique. Due to the interaction between the graphene flake and the atomically flat GaAs surface, the graphene is well attached to the substrate and can be structured mechanically. Scanning the tip of an AFM in contact mode with a high contact force across the surface results in a part being torn out of the flake. The tip hooks behind the flake and pulls it into the direction in which the tip is moving (see Figs. 1a and 1b) because the flake adheres strongly to the surface, it will start to tear along the tip's path. The place where it starts to tear mainly depends on the weakest point near the path of the tip, resulting in rather wide pieces of graphene (up to 1 μm) being torn away. These experiments show that it is indeed possible to displace and even tear apart a graphene flake mechanically; however they also show that controlled nano-machining remains rather difficult. The use of sharp diamond-coated tips [30, 31] may improve these, as yet rather crude, scratching techniques considerably.

Nevertheless, it still possible to peel off individual graphene layers from a flake positioned on GaAs (see Fig. 2); The AFM tip hooks behind the upper layers and peels them off the lower layer(s). Although a very delicate process it opens the possibility to create single layers on GaAs by AFM rather than going through the tiresome procedure of locating one.

An alternative and indeed most promising technique to manipulate a surface with the AFM is local anodic oxidation [18] (LAO). By applying a bias voltage between the AFM tip and the substrate in a humid environment, the substrate directly beneath the tip is oxidized. Controlling the applied voltage and the position of the tip provides ample control to create any desired surface morphology and structure.

For an application of this technique we used fully contacted single-layer and double layer graphene devices deposited on a 300 nm thick $SiO_2$ layer on top of heavily doped Si substrate. These contacts might later be used as device contacts when deposited at suitable positions. We placed the devices under an AFM in an environment with a controlled humidity (55 – 60 %) which allows the formation of a water meniscus between the AFM-tip and the device surface; a schematic setup is shown in Fig. 3. By applying a positive voltage between the graphene sheet and the tip, the graphene can be locally oxidized below the tip following the concept of electrochemical oxidation. At the cathode the current-induced oxidation of carbon leads to the formation of a variety of carbon-based oxides and acids that will escape from the surface and a groove forms in the graphene sheet directly underneath the AFM tip.

Figure 4 shows the experimental realization of this principle: The (doped) silicon tip of the AFM is moved in contact mode slowly ($v_{tip}$ = 0.05 µm/s) across a contacted few-layer graphene flake with a constant voltage ($V_{ox}$ = 25 V) applied between the tip and the graphene sheet. During this process the environment is kept at a constant humidity of 55 % at a temperature of 27 °C. As the graphene flake is oxidized in half, the resistance measured across the flake drastically increases (Fig. 4c). Figure 4a shows an AFM micrograph of the resulting groove with a width of less than 30 nm, as can be seen in the cross section of Fig. 4b along the indicated line in Fig 4a. The remaining graphene on both sides of the groove stays intact, making them ideal for graphene in-plane gates [15, 32] in more complicated structures. The width of the oxidized grooves typically varies between 30 and 100 nm, mainly depending on the apex of the used AFM-tip, and thereby defines the limit on the resolution possible with this technique.

Although the principle of local anodic oxidation sound rather straightforward, it is important to remark that this technique only works if the line is stared at the edge of a graphene sheet. Oxidizing bulk graphite or starting the oxidation in the middle of a graphene sheet turned out to be practically impossible even with voltages up to 40 V. Most likely the carbon-carbon bonds in the center of a graphene sheet are too strong to be broken directly. In contrast, the edge-termination of graphene [33] by e.g. hydrogen atoms can substantially facilitate the initial oxidation process. Additionally, the hydrophobic character of graphite will repel water necessary to from a meniscus between tip and substrate during the oxidation procedure. This hydrophilic behavior of a graphene surface is clearly visible in Fig. 5, where the high environmental humidity (58%) leads to the formation of water droplets, (indicated by the circle) on top of a single layer graphene sheet [34]. These droplets only form on the hydrophobic graphene, but not on the more hydrophilic $SiO_2$ substrate where rather a homogeneous wetting by a water film takes place. As a consequence, we can observe a stripe of water along the edges of the graphene sheet (dotted line in Fig 5, see also Fig. 4) which will substantially ease oxidation from the edges.

## 3. Conclusions

In summary, we have shown that it is possible to use an AFM to nano-manipulate individual layers of graphene either by means of mechanical peeling or by electrochemical oxidation. Local anodic oxidation proved to be an extremely useful manner to manipulate graphene. By oxidizing grooves of less than 30 nm wide in a graphene sheet it is in principle possible to cut out every structure imaginable, e.g. quantum point contacts or quantum dots, making this technique very promising for table top graphene based devise fabrication.

**Figures**

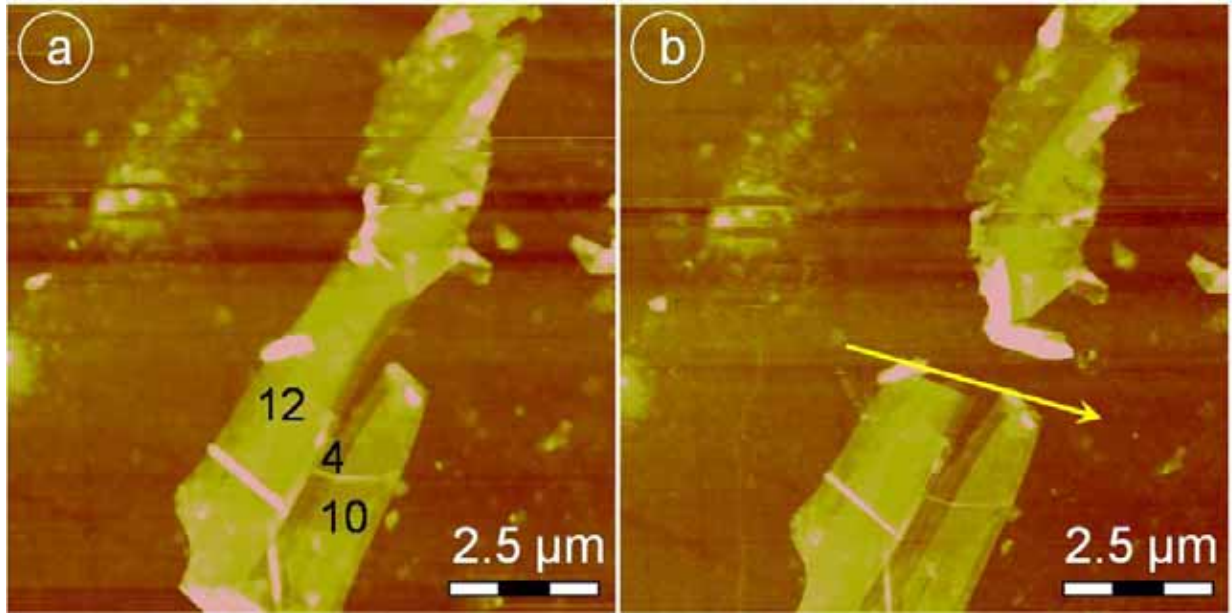

Fig. 1. (Color online) Brute-force mechanical manipulation of few-layer graphene flakes on GaAs ((a) and (b)). The number of layers is depicted in the figure. The flake in (a) is approached from the left with the AFM tip. The flake rips apart and rolls up to the top. Due to the strong van-der-Waals interaction between the graphene and the GaAs substrate, the remainder of the flake remains sticking to the GaAs.

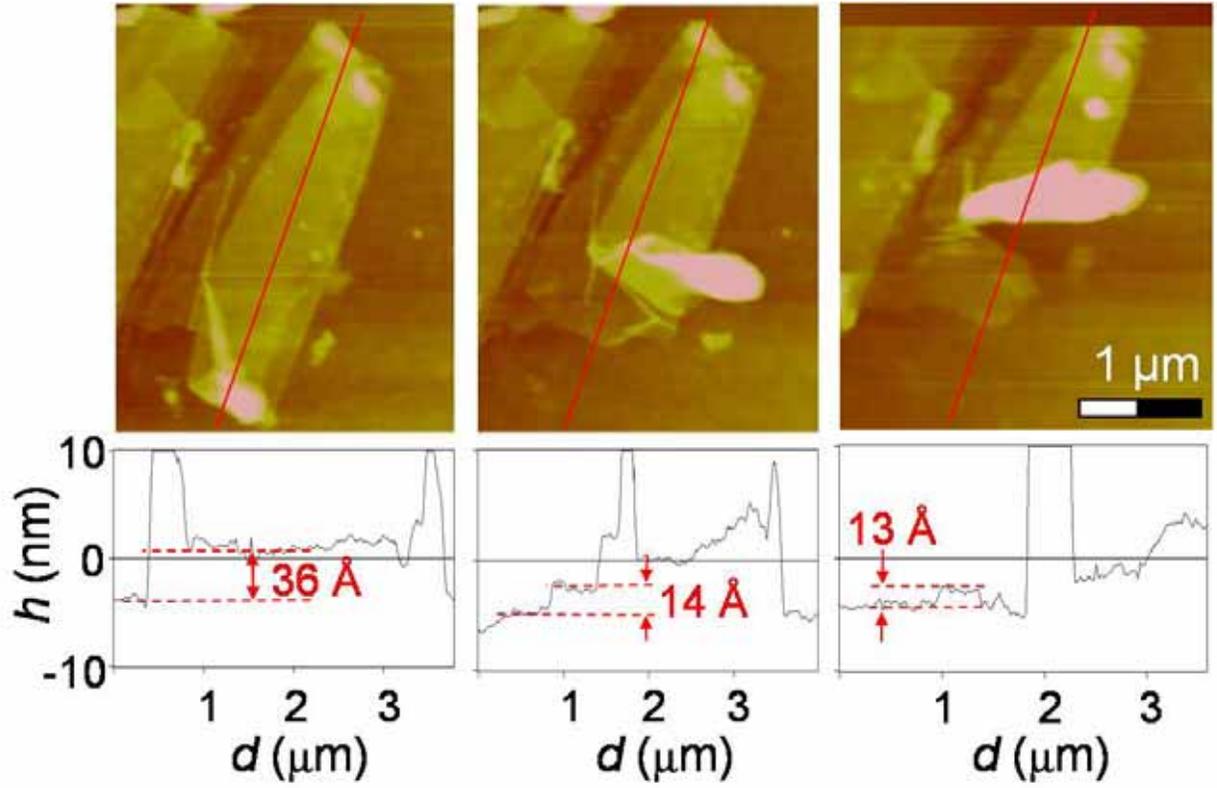

Fig. 2. (Color online) AFM micrographs after three subsequent nano-peeling steps (left to right) of a few layer graphene flake. The pictures on the bottom show the height profiles of each successive step along the lines indicated in the micrographs. The number of layers in the indicated area is reduced from eight to two, leaving an electronically much more interesting graphene bilayer.

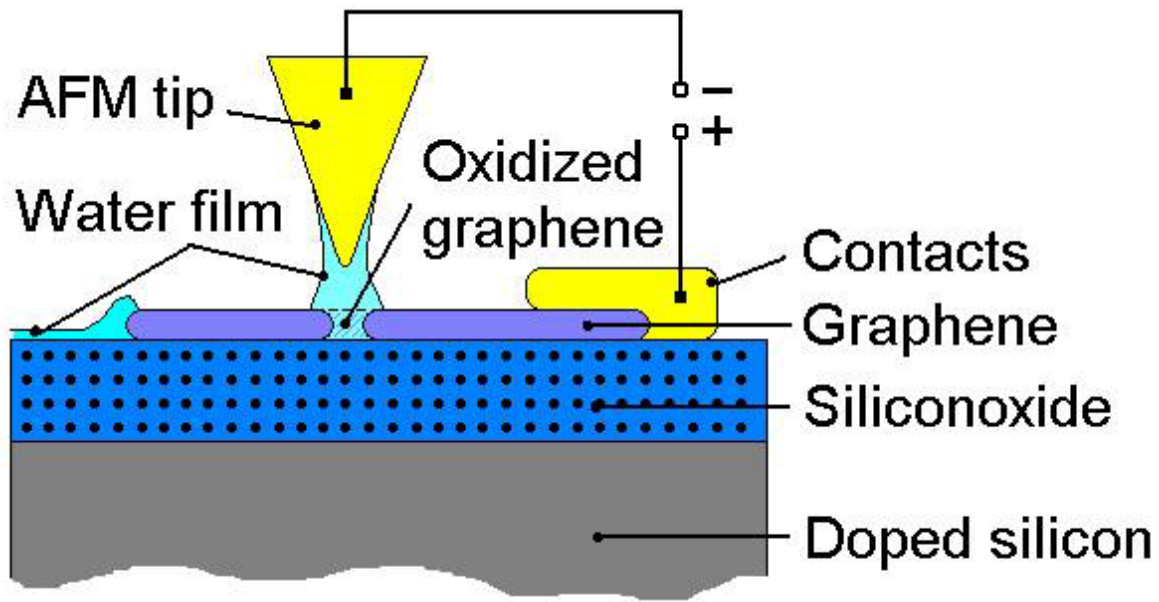

Fig. 3. (Color online) Schematic setup for the local anodic oxidation of graphene. A graphene sheet lies on a SIMOX-substrate and is electrically connected by Au electrodes. A positive bias voltage is applied to the graphene sheet (anode) with the tip of the AFM (cathode) grounded. In a humid environment a water meniscus forms between the AFM and the graphene flake which acts as an electrolyte.

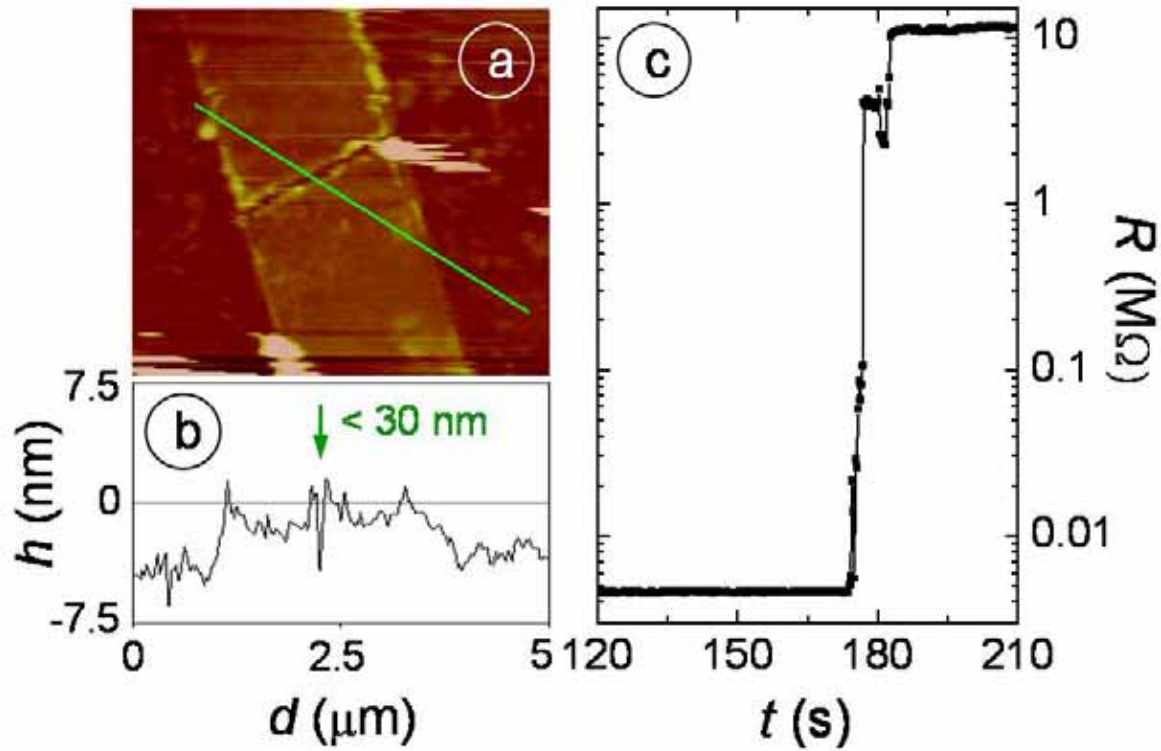

Fig. 4. (Color online) Resistance measurement during the oxidation of a six-layer graphene ribbon. (a) AFM-micrograph taken directly after the oxidation with an unbiased tip. It nicely shows the groove with a line-width of less than 30 nm, where the carbon atoms are removed. (b) Depicts a cross-section of the few-layer graphene ribbon along the line as indicated in (a) showing that the ribbon is clearly oxidized in half. (c) The resistance measured across the ribbon during oxidation increases dramatically as the ribbon is oxidized into two separate parts.

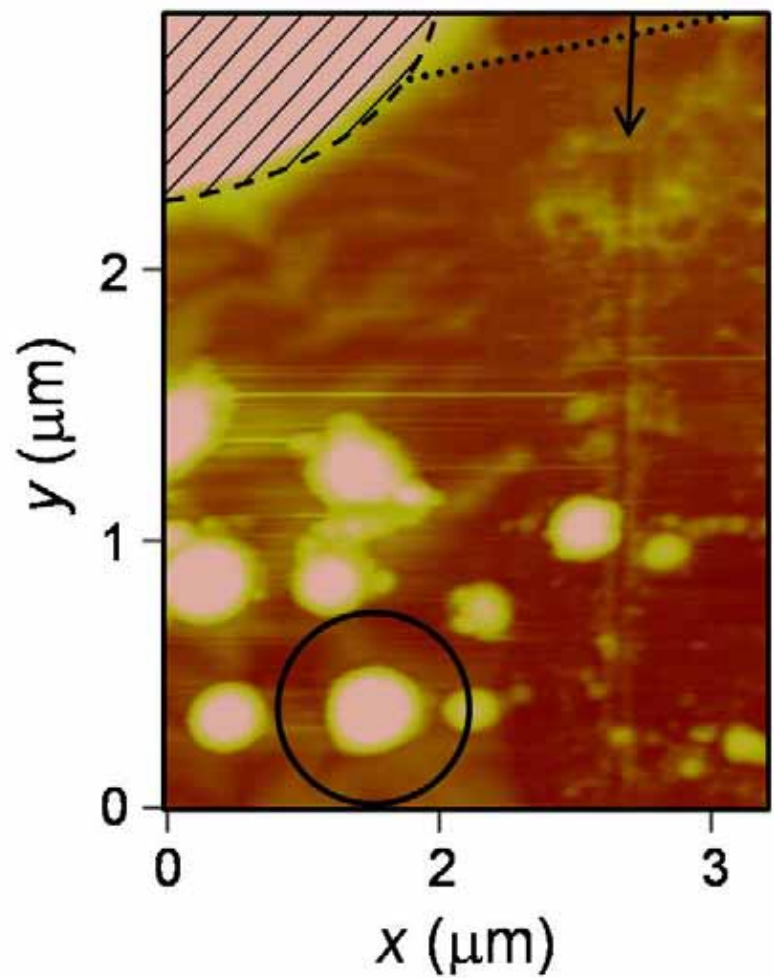

Fig. 5. (Color online) Atomic force micrograph of an oxidized line in a single layer graphene flake. The vertically oxidized line (see arrow) is started at the edge of the flake indicated by the dotted line. Clearly visible are the water droplets (one is encircled) formed on the graphene surface due to the high humidity and hydrophobic character of the graphene. The dashed region is one of the gold contacts to the graphene sheet that serves as cathode during the oxidation procedure.